\documentclass{aa}
\usepackage{graphicx}
\usepackage{amsmath}
\usepackage{natbib}
\bibpunct{(}{)}{;}{a}{}{,}
\begin{document}

\title{The thermohaline, Richardson, Rayleigh-Taylor, Solberg--H{\o}iland and GSF criteria
 in rotating stars}
\titlerunning{Hydrodynamic instabilities}

\author{Andr\'e Maeder$^1$, Georges Meynet$^1$, Nad\`ege Lagarde$^2$ and Corinne Charbonnel$^{1,3}$ }
\authorrunning{Maeder et al.}

\institute{$^1$ Geneva Observatory, Geneva University, CH--1290 Sauverny, Switzerland,\\ $^2$ School of Physics and Astronomy,
University of Birmingham, Edgbaston, Birmingham B15 2TT, UK,\\$^3$ IRAP, UMR 5277 CNRS and Universit\'e de Toulouse,  14, Av. E.Belin, 31400 Toulouse, France\\
              email: andre.maeder@unige, georges.meynet@unige.ch, lagarde@bison.ph.bham.ac.uk, corinne.charbonnel@unige.ch
}

\date{Received  / Accepted }

\offprints{Andr\'e Maeder} 
   
\abstract{}{We examine the interactions of various instabilities in rotating stars, which usually are considered as independent.}{An analytical study of the problem is performed, account is given to radiative losses, $\mu$--gradients and horizontal turbulence.}{
The  diffusion coefficient for an ensemble of instabilities is not given by the sum of the specific coefficients for each instability, but by the solution of a general equation. We find that thermohaline mixing is possible in low-mass red giants only if the horizontal turbulence is very weak. In  rotating stars the Rayleigh--Taylor  
and the shear instabilities need  simultaneous treating. This has for consequence that rotation laws of the form $1/r^{\alpha}$ are predicted to be unstable for $\alpha > 1.6568$, while the usual Rayleigh criterion predicts instability only  for 
$\alpha > 2$.  Also, the shear instabilities are somehow reduced in Main Sequence stars by the effect of the
Rayleigh--Taylor criterion. Various instability 
criteria should be expressed differently in rotating stars than in simplified  geometries.}{}


\maketitle

\section{Introduction}

Various instabilities and processes  intervene in stars, particularly in rotating stars, and produce some
mixing in  radiative zones. These processes are often expressed by diffusion coefficients 
\citep{EndalS78,Zahn92}.
Currently the global diffusion coefficient is expressed as the sum of the  particular 
coefficients of the various effects considered independently. However, there are many physical interactions between the different instabilities, with possible amplification or damping effects. Our aim here is to consider  the interaction
of different instabilities  and their associated criteria. 

The effects  and criteria we consider here are the thermohaline instability  \citep{Ulrich72,KippenhahnRT80},
the shear instability expressed by the Richardson criterion, the Rayleigh--Taylor  instability   \citep{Chandrasekhar61}, the  semiconvective instability, the instability  characterized by the Solberg-H{\o}iland criterion \citep{Wasiutynski46} and the baroclinic instability  expressed by the Golreich--Schubert--Fricke criterion \citep{GoldreichS67,Fricke68}. Account needs to be given also
to non--adiabatic effects, to the $\mu$--gradients and to the horizontal turbulence generally present in differentially rotating stars.
The interactions of horizontal turbulence with shear instabilities  and meridional circulation were studied respectively by \citet{TZ97} and by \citet{ChaboyerZ92}. The exact amplitude of the horizontal turbulence is uncertain and various estimates have been performed \citep{Zahn92,Maeder03Dh,Mathis04Dh}.

An interesting consequence of this work is that some of the well known instabilities
should never be considered \emph{alone} in rotating stars, but always in their interactions with
 other ones. This leads to modified expressions of some well known criteria when applied to rotating stars.

In Sect. 2, we start by deriving  the basic expression of thermohaline mixing in two ways, both showing that a proper account of the mean molecular weight is often missing in many derivations.  
We also examine analytically the effect of the horizontal turbulence on thermohaline mixing. In Sect. 3,
 we show that many instabilities are interacting and should not  be treated separately, but are better analysed in the frame of a more general equation. In Sect. 4, we  reconsider some well known stability criteria and provide new expressions for them in rotating stars. Sect. 5 gives the conclusions.
 
 \section{The double - diffusive mixing  revisited}

 \subsection{The thermohaline mixing}

 We first consider the effects of
 the gradients of mean molecular weight $\mu$ on the stability, since such gradients are always present in stellar interiors as a result of nuclear reactions and ionization. Most of the time the $\mu$--gradients have a stabilizing effect, since generally 
  regions with higher $\mu$ are deeper in the star.
  
 The $\mu$--gradient may also be destabilizing, when  in a medium 
subject to gravity  some upper layers have a higher mean molecular weight  $\mu$ than deeper layers ($\nabla \mu <0$). The medium
may nevertheless be stable with respect to the Ledoux criterion for convection.  This is possible, if   the upper layers with higher $\mu$ are hot enough to allow a stable density gradient. However, if  heat diffusion is fast (faster than  matter diffusion), the medium of higher $\mu$ has the time to cool,    
become denser  and start sinking. This  is the thermohaline instability, which takes  the form of "salt fingers" in lab experiments. 

The interest for thermohaline mixing has been revived over
recent years in relation with the observed chemical abundances of red giants, which require some significant internal mixing at the level of the H-burning shell.
\citet{CZ07} and \citet{CL10} investigated 
the effects of thermohaline instability in red giant stars using a 
classical treatment for the associated diffusion coefficient; they found 
in particular an excellent agreement between theoretical predictions and 
the observations of Li, C, and N isotopes on the upper red giant branch. 
Also,  \citet{Lagarde12} showed that the inclusion of 
thermohaline mixing in stellar models provides a very elegant solution 
to the long-standing helium-3 problem in the Galaxy. On the other side, \citet{CantielloLanger2010} and \citet{Wachlin2011} showed  that to explain the surface abundances observed at the bump in the luminosity function, the speed of the thermohaline mixing process needs to be more some orders of magnitude higher than in their models. These different results are due, at least for a large part, to the different geometries adopted  for  the fluid elements. In these last two works, the fluid elements are
essentially spherical, while other authors like \citet{CL10} in the line of \citet{Ulrich72} consider elongated ``salt fingers''. We will see below (Eq. \ref{Dmu}
and \ref{Dmumu}) that the ratio of the vertical length of the salt fingers to their  diameter intervene to the square as a multiplicative factor in
the thermohaline diffusion coefficient.

The theory has been analyzed by  \citet{DenissenkovP08} in relation with $^3$He burning in the envelopes of red giants, a work  confirmed by 2D numerical simulations \citep{D10}. 
Further numerical simulations \citep{DM11}
suggest that the thermohaline mixing is  not efficient enough
(by  a factor of 50) to account for the observed composition anomalies of red giants. Numerical 3D simulations have been  performed \citep{Traxler11}, these authors develop new prescriptions
and support the inability of thermohaline mixing to achieve the observed mixing, while \citet{Brown2012} find more 
transport than suggested by Traxler et al.
The effect of radiative levitation and turbulence on thermohaline mixing are nicely analyzed by  
 \citet{VauclairT12}. On the whole, we see a large variety of results, often related to the choice of some physical parameters.

 \subsection{The account of the chemically inhomogeneous surroundings}  \label{exnum}

 Often a perturbation method is used to derive  the  coefficient of thermohaline diffusion in the line of \citet{Ulrich72} or a blob method following \citet{KippenhahnRT80}. The problem is that the possible ambient variations of $\mu$ are generally neglected. 
 Let us  consider  the blob method. \citet{KippenhahnRT80}
 have shown  that the falling velocity $v_\mu$ of a blob with an excess
 $\Delta \mu > 0$ with respect to the surrounding medium  is  given by

  \begin{eqnarray}  
v_{\mu} \, = \, \frac{- H_P}{(\nabla_{\mathrm{ad}}-\nabla) \, \tau_{\mathrm{therm}}} \,
\frac{\varphi}{\delta} \, \frac{\Delta \mu}{\mu}  \; ,
\label{vmu}
\end{eqnarray}

\noindent
where $\tau_{\mathrm{therm}}$  is the thermal adjustment timescale  given by 
 \begin{eqnarray}
  \tau_{\mathrm{therm}} = \frac{C_P \, \varrho^2 \,\kappa \,  d^2}{8 \, a \, c \, T^3} \, =  \,\frac{d^2}{6 \, K}  \; ,
  \label{tempstherm}
  \end{eqnarray}
  
  \noindent
  where $K$ is the thermal diffusivity $ K \, = \, {4 \, a \, c\, T^3}/({3 \, \kappa \, \varrho^2 C_{\mathrm{P}}})$ and $d$ the diameter of the blob. The factor of 6 applies to a spherical bubble, while elongated blobs like salt fingers have larger values, e.g. 12, 20 or more.   
The usual notations are used, in particular $H_{P}$ the pressure scale height, $\delta = -
  ({\partial \ln \varrho}/{\partial \ln T})_{P, \mu} \; ,  \;  \; \varphi= 
   \left({\partial \ln \varrho}/{\partial \ln \mu}\right)_{P,T}$.
In a radiative   medium $\nabla_{\mathrm{ad}} > \nabla$, thus for
$\Delta \mu >0$ one has $v_{\mu}<0$ and  the blob is sinking.

The thermohaline diffusion coefficient is $D_{\mathrm{thl}} = \left|\ell \, v_\mu\right|$ (because the transport is anisotropic, there is no factor $1/3$),
thus one has
\begin{eqnarray}
D_{\mathrm{thl}} \,= \, 6 \, \left(\frac{\ell}{d}\right)^2 \, K \, \frac{\varphi}{\delta} \; \,
\frac{(-\nabla_\mu)}{\nabla_{\mathrm{ad}} -\nabla} \; ,
\label{Dmu}
\end{eqnarray}

\noindent
There, for the expression of the relative excess of mean molecular weight, the following expression is generally  $\Delta \mu /\mu= \ell \, (d \ln \mu/dr)= - \ell \, \nabla \mu \,/ H_P$, where 
$\ell$ is the mean free path of the fluid element. 
The geometrical factors  $6 \, \left({\ell}/{d}\right)^2$ can be  grouped in a constant $C_{\mathrm{thl}}$, which may become large for salt fingers. An expression of the 
form (\ref{Dmu}) has been found by \citet{Ulrich72} with a coefficient $C_{\mathrm{thl}}=685$, a similar expression but with  $C_{\mathrm{thl}}=12$ was obtained by \citet{KippenhahnRT80}. The same expression but with $C_{\mathrm{thl}}=1000$
has been used by \citet{CL10}, implying that the mixing-length of the fluid elements is very large with respect their diameters. Similar expressions, but with different coefficients, were also obtained and used by \citet{DenissenkovP08}, \citet{TheadoV12} and \citet{VauclairT12}. 

A critical point is that expression (\ref{vmu}) and consequently (\ref{Dmu})  apply for thermohaline mixing  in  a surrounding \emph{homogeneous} medium. This is  not  the general case, e.g. in a radiative shell H-burning region,  there is a
$\mu$--gradient throughout the zone where mixing develops. (We may also remark that in the derivation of (\ref{Dmu}) it is a bit strange  to  assume the surrounding medium homogenous and at the same time to make use of a certain $\nabla \mu$. Maybe,  this $\nabla \mu$ could be considered as the average gradient between the blob and the surrounding medium, but this is not very satisfactory). 

Thus, let us consider an inhomogenous medium with a certain $\mu$--gradient, the  general form of the sinking velocity $v_{\mu}$ is thus \citep{maederlivre09},

\begin{eqnarray}  
v_{\mu} \, = \, \frac{- H_P}{\left(\nabla_{\mathrm{ad}}-\nabla+\frac{\varphi}{\delta} \, \nabla_{\mu}\right) \, \tau_{\mathrm{therm}}} \, \frac{\varphi}{\delta} \, \frac{\Delta \mu}{\mu}  \; .
\label{vmumu}
\end{eqnarray}

\noindent
As a consequence, the thermohaline diffusion coefficient is
\begin{eqnarray}
D_{\mathrm{thl}} \, = \, 6 \, \left(\frac{\ell}{d}\right)^2 \, K \, \frac{\varphi}{\delta} \; \,
\frac{(-\nabla_\mu)}{\nabla_{\mathrm{ad}} -\nabla +\frac{\varphi}{\delta} \, \nabla_{\mu}} \; .
\label{Dmumu}
\end{eqnarray} 

\noindent
This expression supports the consistent finding by \citet{D10} of a term $\nabla \mu$ at the denominator of (\ref{Dmumu}), a term which is generally ignored, both in theoretical papers and modeling of stellar evolution. 
We examine the size  of the various terms in the numerical model of a 1.25 $M_{\odot}$ of solar composition, with an initial rotation velocity of 110 km s$^{-1}$ at the stage of the red giant bump with a luminosity of 105 $L_{\odot}$, as in Fig. 9 by  \citet{CL10}.  The zone where thermohaline mixing is present extends over about 2.6 \% of the radius around a mass fraction
$M_R/M = 0.30$ at the base of the extended outer convective zone. Over most of this small zone,
$(\nabla_{\mathrm{ad}} -\nabla)$ is around
0.15, while  the term $(\varphi/\delta) \, \nabla_{\mu}$ is about $3 \cdot 10^{-6}$, meaning that, even if $\nabla_{\mu}$ is the leading term at the numerator its effect at the denominator is  negligible here.  As a result the coefficient of thermohaline 
diffusion is high, lying between $10^7$ and  $10^{10}$ cm$^2$ s$^{-1}$, while the  coefficients resulting from other diffusion sources are in the range of $10^2$ to $10^3$ cm$^2$ s$^{-1}$. 
So far so good, this would  confirm the importance of the thermohaline mixing,  however the thermohaline instability, like other instabilities, is also influenced 
by non--adiabatic effects and turbulence, which we now consider.

\subsection{The effects of thermal losses and turbulence }  \label{thalother}

 Let us first consider the effects of radiative thermal losses.
  In a medium with gravity,
the net acceleration (by unit of length) due to the  density difference between a fluid element 
 and the surroundings is, 
\begin{eqnarray}
 g \, \frac{d(\delta \ln \varrho)}{dr} =   \frac{g}{H_P}\left[\delta(\nabla_{\mathrm{int}} - \nabla)
- \varphi (\nabla_{\mathrm{int,\mu}} - \nabla_{\mu} \right)] \, , 
\label{dzrho2}
\end{eqnarray}
\noindent
where the two parentheses on the right contain the differences between the internal and external gradients. The difference of $T$--gradients depends on the radiative losses determined by the thermal diffusivity $K$ \citep{Maeder95},
 \begin{eqnarray}
 \nabla_{\mathrm{int}} - \nabla \, = \, \frac{\Gamma}{\Gamma+1} 
 \left(\nabla_{\mathrm{ad}} -\nabla \right) \; \; \; 
 \mathrm{and} \quad  
 \Gamma = \frac{v \, \ell}{6 \, K} \, ,
 \label{G+11}
 \end{eqnarray}
\noindent
where $\Gamma$ is the ratio of the energy transported to the energy lost on the way (i.e. $\Gamma$ equals the Peclet number over 6), with $v$ and $\ell$ the typical velocity and mean free path of the  motions. \\

Let us now turn to turbulence, which originates mainly from differential rotation, both vertical (producing vertical shears) and horizontal (producing horizontal shears creating horizontal turbulence).

The (vertical) shear instability is produced by  a significant  difference $\Delta v$ of velocity between nearby layers of different vertical coordinate $r$. In a shear, there is an  excess  energy by unit of mass and surface,
    \begin{equation}
  \frac{1}{4} \;  \left(\frac{\delta v}{\delta r} \right)^2 \; ,
  \label{dv}
  \end{equation}
  \noindent
   locally present in the differential motions
 with respect to the case of  an average velocity. The Richardson criterion  defines the occurrence of shear instability over the zone in $r$ considered.
 If the excess energy in the shear is greater than the work necessary for overcoming
the stable temperature and $\mu$--gradients, we get the usual Richardson criterion (see also Sect. \ref{sectri}).

As to the horizontal turbulence, it intervenes by changing the $\mu$--gradients.
 \citet{TZ97} have shown that the coefficient
of horizontal turbulence $D_{\mathrm{h}}$ plays the same role
as the thermal diffusivity in the previous expression (\ref{G+11}) and  write,
\begin{eqnarray}
\nabla_{\mu, \, \mathrm{int}}- \nabla_{\mu} \, = \, -\frac{\Gamma_{\mu}}{\Gamma_{\mu}+ 1} \nabla_{\mu}
\quad \mathrm{with} \;\; \Gamma_{\mu }= \frac{v \ell}{6 \, D_{\mathrm{h}}} \; .\\ \nonumber
\end{eqnarray}

\noindent
With proper account of heat losses and horizontal turbulence,  the modified
 Richardson criterion for shear instability  becomes \citep{TZ97},
\begin{eqnarray}
\left(\frac{\Gamma}{\Gamma+1}\right) N^2_{ \mathrm{ad}} + \left(\frac{\Gamma_{\mu}}{\Gamma_{\mu}+1}\right) N^2_{\mu}
\, < \, \mathcal{R}i_{\mathrm{c}} \left(\frac{dv}{dr}\right)^2 \; ,
 \label{premg} 
 \end{eqnarray}
 \begin {eqnarray}
\mathrm{with} \quad   N ^2_{\mathrm{ad}} = \frac{g  \delta}{H_P} \left(
 \nabla_{\mathrm{ad}}-\nabla_{\mathrm{}} \right), \quad \mathrm{and}\quad N^2_{\mu} = \frac{g \, \varphi}{H_P} \nabla_{\mu} \,.
 \label{rimu}
 \end{eqnarray}
\noindent
The standard value of $\mathcal{R}i_{\mathrm{c}}=1/4$.
Following \citet{TZ97}, we write with $x= v \, \ell/6$,
\begin{eqnarray}
\frac{x}{x+K+D_{\mathrm{h}}} \, N^2_{\mathrm{ad}} + \frac{x}{x+D_{\mathrm{h}}} \,  N^2_{\mu} \, < \, \mathcal{R}i_{\mathrm{c}}
\left(\frac{dv}{dr}\right)^2 \; .
\label{xdh}
\end{eqnarray}
\noindent
The heat transport by $D_{\mathrm{h}}$ adds its effects to $K$. 
Assuming that the shear diffusion coefficient $D_{\mathrm{shear}}= (1/3) v \, \ell =2x$  (the transport concerns the vertical direction)  is small with respect to $K$ and $D_{\mathrm{h}}$, \citet{TZ97}
have obtained $D_{\mathrm{shear}}$ with account of horizontal turbulence and heat losses. 
Let us note that this applies to a spherical fluid element. If not, the result for $x$ has to be 
multiplied by $\left({\ell}/{d}\right)^2$ as in (\ref{Dmu}). 

\subsection{Limit for thermohaline instability}       

If there is no energy available from the differential rotation for shear instability, the second member of (\ref{xdh}) is zero. Then, if we also ignore the horizontal turbulence, we just get the limit
\begin{eqnarray}
D_{\mathrm{thl}} \, = \, 6 \, \left(\frac{\ell}{d}\right)^2 \, \left(\frac{ - K \, N^2_\mu}{N^2_{ \mathrm{ad}}+N^2_\mu} \right) \; . 
\end{eqnarray} 
\noindent
This is the same as ($\ref{Dmu}$), it also shows the presence of a term depending 
on the $\mu$--gradient at the denominator. For thermohaline mixing,
$\nabla \mu$ and $N^2_\mu$ are negative, thus the diffusion coefficient is evidently positive.
As mentioned above, $N^2_\mu$ at the denominator increases the mixing.\\

If some horizontal turbulence is present, but no significant shears, we get from (\ref{xdh})

\begin{eqnarray}
D_{\mathrm{thl}} \, = \, 6 \, \left(\frac{\ell}{d}\right)^2 \, \left(\frac{ - K \, N^2_\mu}{N^2_{ \mathrm{ad}}+N^2_\mu} \;- \, D_{\mathrm{h}} \right) \; 
\label{ddd}. 
\end{eqnarray}
\noindent
The first term in the parenthesis is positive and the second negative. Thus, for mixing to occur, one must have a limit on the value of the horizontal turbulence $D_{\mathrm{h}}$, 
\begin{eqnarray}
D_{\mathrm{h}} \; < \; K \, \frac{ - N^2_\mu}{N^2_{ \mathrm{ad}}+N^2_\mu} \; .
\label{comp}
\end{eqnarray}
\noindent
A similar expression, but without the term in $N^2_\mu$ at the denominator, has been obtained  
by \citet{DenissenkovP08}. 
In general, the numerical calculations show that $D_{\mathrm{h}}$ is smaller than $K$. However,
for thermohaline mixing to occur, it is necessary that at the beginning (before the slow descending motions) the system
is stable, thus $N^2_{\mathrm{ad}} >> \left|N^2_{\mu}\right|$, this means that the r.h.s. of (\ref{comp}) may be much smaller than $K$.
Thus, $D_{\mathrm{h }}$ becomes a factor which may cancel the large parenthesis in (\ref{ddd}) and stop the mixing. This is consistent: when the differences of $\mu$ are small with respect to the stabilizing effect of the $T$-- gradient,
turbulence may easily kill the thermohaline mixing.

As a numerical example, we consider the 1.25 $M_{\odot}$ model discussed in Sect. \ref{exnum}. In this example, the r.h.s of (\ref{comp}) is  1-2 orders of magnitude larger than 
$D_{\mathrm{h}}$, implying that the adopted horizontal turbulence does  not inhibit the thermohaline mixing. 
However,  the situation may not be so simple. As mentioned, there is a great uncertainty in the size of the horizontal turbulence   and 
three different expressions have been proposed \citep{Zahn92,Maeder03Dh,Mathis04Dh}.  \citet{CL10} have taken the lowest bound
of the first one. The last two, likely more physical \citep{maederlivre09}, give similar numerical values, larger by 4 to 5 orders of magnitude than
the first one. If applicable, such high values of $D_{\mathrm{h}}$ are sufficient to kill
the thermohaline mixing. Thus, the problem of the thermohaline mixing remains open and we see that a key to its solution lies in a better knowledge of the horizontal turbulence in stars, in addition to the proper choice of the  geometrical parameters of the fluid elements.

\subsection{Semiconvective diffusion}

We now turn to semiconvection, for which the 
 expression of the diffusion  coefficient  is related to that of the thermohaline mixing.
Formally, semiconvection occurs when the thermal gradient $\nabla$ is intermediate between the condition of stability predicted by the Ledoux criterion and the instability predicted by Schwarzschild's criterion,
\begin{equation}
\nabla_{\mathrm{int}} \, < \, \nabla \, < \, \nabla_{\mathrm{int}}+ \frac{\varphi}{\delta} \, \nabla_{\mu} \; ,
\label{SL}
\end{equation} 
\noindent
$\nabla_{\mathrm{int}}$ being very close to $\nabla_{\mathrm{ad}}$ in stellar interiors.
The velocity of the semiconvective mixing is also given by (\ref{vmumu}) as for thermohaline mixing. 
The blobs being in dynamical equilibrium, the equation of state implies
$\frac{\Delta T}{T} \, = \, \frac{\varphi}{\delta} \, \frac{\Delta \mu}{\mu}$, where
$\Delta T$ for a blob displaced over a distance $\Delta r$ is
$\frac{\Delta T}{T}\, = \, \left(\nabla_{\mathrm{int}} - \nabla \right) \, \frac{\Delta r}{H_{P}}$.
With account of (\ref{tempstherm}) and (\ref{G+11}), a coefficient of semiconvective 
diffusion  may be obtained \citep{maederlivre09},
\begin{eqnarray}
D_{\mathrm{sc}} \, = \, \frac{2 \, \Gamma}{\Gamma+1} \, \,\left| \frac{K \, (\nabla-\nabla_{\mathrm{ad}})}
{\left(\nabla_{\mathrm{ad}} -\nabla + \frac{\varphi}{\delta} \nabla_{\mu} \right)}\right| \; .
\label{dsc}
\end{eqnarray}
\noindent
For $\Gamma$ tending towards infinity (adiabatic case), one gets the coefficient obtained by  \citet{LangerFS83}.  However,
semiconvection arises from nonadiabatic effects and
in stellar interiors, $ \Gamma$ lies between $10^{-2}$ 
 and $10^{-3}$. Thus, the numerical coefficient  $\frac{2 \, \Gamma}{\Gamma+1} $ in front of (\ref{dsc}) is of this order of a magnitude.

\section{Instabilities in rotating stars}  \label{instab}

\subsection{The coupling of the Rayleigh--Taylor and shear instabilities} \label{coupl1}

 The effects of shears, thermal diffusion, thermohaline mixing and horizontal turbulence are all already included in  Eq. (\ref{xdh}). However, we must also account for the stabilizing or destabilizing effects related to the distribution of the specific angular momentum  $j \, = \, \varpi^2 \, \Omega$. Indeed, if a (descending) salt finger brings more specific  angular momentum inwards than present in the surrounding medium, this difference of $j$ will tend to bring the fluid element back to its initial position
after some oscillations of frequency $N_{\Omega}$, thus  the medium is stable. At the opposite, if the salt finger brings less
specific angular momentum than present in the surroundings, it tends to continue its way and an instability is produced. An analogous discussion can be made for outwards motions. This is well  known from the Rayleigh instability criterion: a medium with $j$ decreasing outwards is unstable.

\begin{figure}[t]
\begin{center}
\includegraphics[width=9.0cm, height=6.5cm]{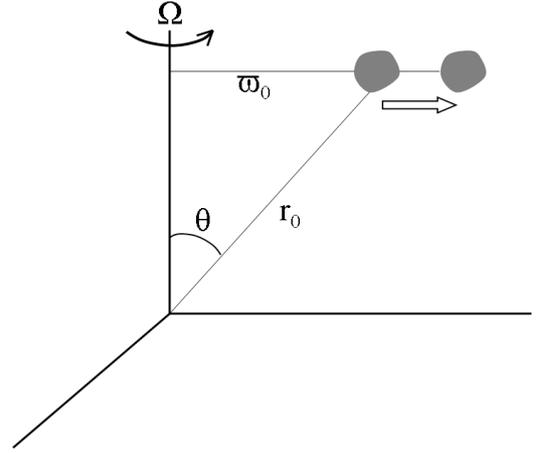}
\caption{Geometry data concerning the Rayleigh--Taylor instability.}
\label{solberg}
\end{center}
\end{figure}

In order to examine  closely this  effect, 
let us consider a star in differential rotation. For now, we need not to specify more the dependence of the angular velocity
$\Omega$ on the spatial variables. Let us consider a fluid element at a
distance $\varpi_0$ and let us move it  outwards, as illustrated in Fig. \ref{solberg}, to a very close point at distance  $\varpi$. The difference  of the centrifugal forces acting on the blob by unit of mass may be written            
\begin{eqnarray}
\varpi \left(\Omega^2_{\mathrm{int}}-\Omega^2_{\mathrm{ext}} \right)=
\frac{1}{\varpi^3}\left(j^2_{\mathrm{int}} -j^2_{\mathrm{ext}}\right)= \\ \nonumber
-\frac{1}{\varpi^3} \, \frac{d(\Omega^2 \varpi^4)}{d\varpi}\, (\varpi - \varpi_0) \, \; .
\end{eqnarray}
\noindent
The work $\delta \mathcal{T}_{\mathrm{centrif}}$ delivered by the bubble to the medium during 
this small displacement is
\begin{eqnarray}
\delta \mathcal{T}_{\mathrm{centrif}}  =  -\frac{1}{\varpi^3} \, \frac{d(\Omega^2 \varpi^4)}{d\varpi} \, (\varpi -\varpi_0)^2 =  \\ \nonumber
- \, N^2_{\Omega} \, \, \delta \varpi^2 \,   . 
\label{workcentrif}
\end{eqnarray}
\noindent
If $\Omega$ decreases very fast with increasing $\varpi$, $N^2_{\Omega}$ may be negative. This means that the motion of the bubble spontaneously delivers some work $\delta \mathcal{T}_{\mathrm{centrif}}$ to the medium, which  is thus unstable. At the opposite, if $N^2_{\Omega}$ is positive, one has to give some energy to the bubble in order to move it outwards. The medium is stable. This leads  to the  Rayleigh  criterion for the stability of a distribution of angular momentum,

\begin{eqnarray}
N^2_{\Omega} = \frac{1}{\varpi^3} \, \frac{d\left( \Omega^2 \, \varpi^4 \right) }{d\varpi} \; > \; 0 \; .
\label{travcentrif}
\end{eqnarray}

We emphasize that the above effects of the work made by the centrifugal force leading to Rayleigh criterion  (\ref{travcentrif}) and the effects 
of shears as expressed by (\ref{dv})  are  two different physical effects. In some particular hydrodynamic situations, these two effects can be treated separately. For example, in plane parallel
flows with different velocities, the problem of centrifugal force is not relevant. At the opposite, \emph{in a differentially rotating star, both shear effects and angular momentum effects may influence the stability and should then be treated simultaneously}.\\

 The  displacement of a fluid element implies the exchange of bubbles,  leading to some mixed fluid element with an average spatial velocity.
As recalled above, this makes an energy 
$(1/4) \, (\delta v)^2$ available by unit of mass. In a star with differential rotation of the "shellular" form
$\Omega(r)$, the velocity difference $\delta v$ is   
\begin{eqnarray}
\delta v \, = \, r \,\sin \vartheta \, \frac{d \Omega}{dr} \, \delta r \; ,
\end{eqnarray}
\noindent
where $\vartheta$ is the colatitude of the point considered (cf. Fig. \ref{solberg}).
Thus,  an energy $\delta \mathcal{T}_{\mathrm{shear}}$
is made available during an exchange,
\begin{eqnarray}
\delta \mathcal{T}_{\mathrm{shear}}= \frac{1}{4} \, r^2 \, \sin^2 \vartheta 
\left( \frac{d \Omega}{dr} \right)^2  \delta r^2 \, .
\label{dtshear}
\end{eqnarray}

\noindent
If, this energy available  from the  shear is smaller than the 
energy (\ref{travcentrif}) necessary to destabilize the angular momentum distribution,
 the medium is stable, i.e.  if
\begin{eqnarray}
\delta \mathcal{T}_{\mathrm{shear}}  \; \leq  \; - \delta \mathcal{T}_{\mathrm{centrif}} \, .
\end{eqnarray}
\noindent
In the equatorial regions for a case with $\Omega(r)$, this can be written,
\begin{eqnarray}
 r^2 \, \left( \frac{d \Omega}{dr} \right)^2  \, \leq
 \, 4 \, N^2_{\Omega}  \; ,
 \label{nr}
 \end{eqnarray}
 \noindent
 or in terms of $\Omega$,
 \begin{eqnarray}
 r^2 \, \left( \frac{d \Omega}{dr} \right)^2  \, \leq \, 16 \, \Omega^2 + 8 \, r \, \Omega \, \frac{d\Omega}{dr}.
 \label{omlim}
 \end{eqnarray}
 The equality of the two members of this expression gives a quadratic equation in  $(d \Omega/dr)$, the solution of which is
 \begin{eqnarray}
 \left( \frac{d \Omega}{dr} \right)  =  
   4  \, \frac{\Omega}{r} \, \left(1 \pm \sqrt{2} \right) \; .
   \label{pm}
   \end{eqnarray}
\noindent
To be consistent with a solution in absence of shear, we take the sign minus in the above expression.  This means that in order to be stable, $\Omega(r)$ should not decrease
faster  than a limiting $\Omega$--distribution,
\begin{equation}
 \Omega_0 \, \left( \frac{r}{r_0} \right)^{4(1-\sqrt{2})} \; \sim  \; \frac{1}{r^{1.6568}} \; .
\end{equation} 
\noindent
If $\Omega$ decreases outwards faster than this limit, the medium is unstable.
Rotation laws of the form $1/r^{\alpha}$ with $\alpha$ between 1.6568 and 2.0 are  predicted to be 
stable by the usual Rayleigh criterion, while if we also  account for the excess of energy available in the shear motions they are unstable. It is clear that there is no reason to ignore the excess of energy available in shear motions, which contribute to destabilize the system. Therefore in a differentially rotating star, the Rayleigh criterion must be modified  (see Sect. \ref{modRT}).

Reciprocally, when shear instability is considered in a rotating star, 
the stabilizing or destabilizing effect of the $\Omega$--distribution must also be accounted for.
Below,  the two effects are coupled and  considered simultaneously. This leads to some modifications of well--known criteria.

\subsection{A more general coupling} \label{gensol}

Let us try to be more general and  consider simultaneously various effects which may influence the stability  of the medium in a rotating star with $\Omega(r)$:

\begin{itemize}
\item the effects of thermal gradients,

\item the thermohaline mixing, 

\item the semiconvective diffusion,

\item  the shear mixing to  the local excess of energy in differential rotating layers,

\item the stabilizing or destabilizing effect of the distribution of angular momentum,

\item the radiative losses, 

\item  the transport of heat by the horizontal turbulence,

\item  the element diffusion in the medium  due to the horizontal turbulence.
\end{itemize}

\noindent
Following Eq.(\ref{premg}) and the previous subsection, also taking into account that centrigugal force operates 
orthogonally to the rotation axis, we write for the complete form of the instability expression,

\begin{eqnarray}
\left(\frac{\Gamma}{\Gamma+1}\right) N^2_{\mathrm{ad}} + \left(\frac{\Gamma_{\mu}}{\Gamma_{\mu}+1}\right) N^2_{\mu} +N^2_{\Omega}  \sin \vartheta
 <  \mathcal{R}i_{\mathrm{c}} \left(\frac{dv}{dr}\right)^2  , 
 \label{rimuomega}
 \end{eqnarray}
 
\noindent
where the standard value of  $\mathcal{R}i_{\mathrm{c}}$ is $1/4$ and we have accounted  for the  effect of the distribution of the specific
angular momentum $j$ on the stability.
Making the same transformation  $x= v \, \ell/6$ as in (\ref{xdh}), we get the following instability condition,

\begin{eqnarray}
\frac{x}{x+K+D_{\mathrm{h}}}  N^2_{\mathrm{ad}} + \frac{x}{x+D_{\mathrm{h}}}   N^2_{\mu} +N^2_{\Omega} \sin \vartheta <  \mathcal{R}i_{\mathrm{c}}
\left(\frac{dv}{dr}\right)^2  .
\label{xdhomega}
\end{eqnarray}

\noindent
This general equation may be used to get the total resulting diffusion coefficient $D_{\mathrm{tot}}= 2 x$ of the chemical elements in a star, taking into account the many effects involved.  
The stability condition may be written as a quadratic equation of the form $Ax^2+Bx+C > 0$,

\begin{eqnarray}
\left[N^2_{\mathrm{ad}}+N^2_{\mu}+N^2_{{\Omega}-\delta v}
\right]\, x^2+ \nonumber \\
\left[N^2_{\mathrm{ad}}D_{\mathrm{h}}+N^2_{\mu}(K+D_{\mathrm{h}})+ 
N^2_{{\Omega} -\delta v}
(K+2 D_{\mathrm{h}}) \right] \, x+ \nonumber \\
N^2_{{\Omega}-\delta v}
(D_{\mathrm{h}}K+
D^2_{\mathrm{h}}) \, > \,0 \; ,
\label{2ndeq}
\end{eqnarray}
\noindent
with

\begin{equation}  
N^2_{{\Omega}-\delta v} = \frac{1}{\varpi^3} \, \frac{d\left( \Omega^2 \, \varpi^4 \right) }{d\varpi} \, \sin \vartheta
 - \mathcal{R}i_{\mathrm{c}} \left(\frac{dv}{dr}\right)^2 .
\label{nodv}
\end{equation}
\noindent
Expression of $N^2_{{\Omega}-\delta v}$
 is a modified form of the Rayleigh oscillation frequency, accounting for both  the angular momentum  distribution and the excess of energy in the shear. 
Eqn. (\ref{2ndeq}) is the general equation, which should 
generally be considered in a rotating star to account for the various effects mentioned above.
It contains the Schwarzschild, the Ledoux, the Solberg--H{\o}iland, the Rayleigh--Taylor, the Richardson and the GSF criteria 
with radiative losses and horizontal turbulence all in one single expression.
Simplifications are possible, but they should be justified.

In the first evolutionary 
phases, $N^2_{\Omega}$ is generally likely to have a stabilizing influence since the $\Omega$--gradient is not very steep, this means that the instabilities could be  somehow reduced in the equatorial regions. In the advanced phases after the end of the He-burning
phases, $N^2_{\Omega}$ may be destabilizing over very limited regions\citep{MaederH10}, there equatorial instabilities may be favoured.

At the equator, coordinates $\varpi$ and $r$ coincide, the effect of the centrifugal force is 
maximum, while at the pole this effect disappears and $N^2_{\Omega}$ vanishes.
The distribution of the instability inside the star is no longer spherically symmetric. Indeed, all properties of rotating stars are two dimensional:
the shape, the isobars, the meridional circulation and the various instabilities have a dependence in colatitude $\vartheta$. The usual practice  \citep{Kipp70} is to consider the
equations of structure at an average radius $r_P$, such that the volume of the sphere of radius $r_P$ is equal to the real volume of the oblate rotating star.
This is also valid for differentially rotating star with shellular rotation \citep{MMI}. The radius $r_P$ is equivalent 
to the radius of the oblate star at colatitude $\vartheta$ given by $P_2(\cos \vartheta)=0$, i.e. $\cos^2 \vartheta = 1/3$, which corresponds to $\vartheta =54.7   $ degrees. Thus, in Eq. (\ref{nodv}), we put $\sin \vartheta= \sqrt(2/3)$ and we may thus keep the advantages of the 1D modeling. However, all these procedures may satisfactorily apply for the lower rotation velocities, while  for high velocities, there may be significant  differences with the complex 2D reality of rotating stars.

\subsection{The general solution}

In stellar models, it is often considered that the total diffusion coefficient $D_{\mathrm{tot}}$ is  the sum of several  diffusion coefficients $D_{\mathrm{i}}$, each one corresponding to an   instability $i$,  
 
 \begin{equation}
 D_{\mathrm{tot}} \; = \; \sum_i{D_{i}} \; .
 \label{sumdi}
 \end{equation}
 
 \noindent 
 Such a summation of coefficients $D_i$  is     
 undoubtedly wrong! \emph{The solution of an equation of the second degree like (\ref{2ndeq}) is not the sum of the limits
 of some particular  cases}. The solution is that of the complete equation. This  has for consequence to introduce 
 correlations between  the various effects in the form of products of  terms appearing in the coefficients
  $A$, $B$ and $C$ of the quadratic equation
 (\ref{2ndeq}). These  are
 \begin{eqnarray}
 A= \left[N^2_{\mathrm{ad}}+N^2_{\mu}+N^2_{{\Omega}-\delta v}
\right] \; ,\\  \nonumber
 B= N^2_{\mathrm{ad}}D_{\mathrm{h}}+N^2_{\mu}(K+D_{\mathrm{h}})+ 
N^2_{{\mathrm{\Omega}}
-\delta v}
(K+2 D_{\mathrm{h}}) \; ,\\ \nonumber
C=N^2_{{\Omega}-\delta v}
(D_{\mathrm{h}}K+
D^2_{\mathrm{h}}) \; .
\label{BC}
\end{eqnarray}
\noindent
 The solution of (\ref{2ndeq})  brings coupling of all intervening factors. We notice in particular the cross--products of the heat losses and of the horizontal turbulence with the effects of the thermal, $\mu$-- and $\Omega$--gradients. There are also products of these three gradients and also direct products of heat losses and horizontal turbulence. 
This indicates that the full solution of the diffusion coefficient $D= \left| 2x \right|$ depends on the coupling of the various physical effects considered and is by no means a sum like (\ref{sumdi}).

\section{Some stability criteria reconsidered}

The above results lead us to reconsider some basic stability criteria in differentially rotating star, which indeed  only account for one specific aspect of the instability and ignore other ones, unavoidably present in rotating stars.  In  particular, we see that the complete equation (\ref{2ndeq}) contains the term  $N^2_{{\Omega}-\delta v}$. In agreement with the result of (\ref{coupl1}),  the term  $N^2_{{\Omega}-\delta v}$   should always be considered instead of  $N^2_\Omega$ in the relevant criteria. 

\subsection{A modified Rayleigh--Taylor criterion}  \label{modRT}

The usual Rayleigh--Taylor criterion for stability  of a rotating flow (in absence of gravity) is
\begin{equation}
N^2_{\Omega} \, > \, 0 \; , \\ [2mm]
\mathrm{with} \quad N^2_{\Omega} = \frac{1}{\varpi^3} \, \frac{d\left( \Omega^2 \, \varpi^4 \right) }{d\varpi} \; .
\label{classicRT}
\end{equation}
\noindent
It expresses that in order a rotating medium is stable, the angular momentum must increase with the distance to the rotation axis.

If in the general equation (\ref{2ndeq}) we put $K$ and $D_{\mathrm{h}}$ equal to zero and if we also
do not consider the effects of thermal and $\mu$--gradients,
 we are left with the stability condition 
 
\begin{equation}
N^2_{{\Omega}-\delta v} \, > \, 0  \; .
\label{rtm}
\end{equation} 
\noindent 
At a colatitude $\vartheta$ in a star, we should  consider  this new form of the criterion for the stability of differential rotation,
If positive, it means that the excess energy in the differential rotation is unable to destabilize a stable distribution of angular momentum. If negative, the medium is  unstable.

 Indeed, the Rayleigh--Taylor criterion (\ref{classicRT}) may  predict stability
of a given differential rotation law, while in fact it would  be unstable according to (\ref{rtm}) as discussed in Sect. \ref{coupl1}.  The usual Rayleigh--Taylor criterion
 (\ref{classicRT}) should  never apply as such in a rotating star,
since the shears also favour instability.

 \subsection{A modified form of the Solberg--H{\o}iland criterion} \label{solb}
 
Let us first recall the usual Solberg--H{\o}iland criterion for convective stability,
\begin{equation}
 N^2_{\mathrm{ad}}+N^2_{\mu}+N^2_{\Omega} \, \sin \vartheta \, > \, 0 \; . 
 \label{solbh}
\end{equation}

\noindent 
This to some extent combines   the Ledoux and the  Rayleigh-Taylor criteria.
It expresses the stability with respect to convective motions (at the dynamical timescale
$R^3/(GM)$) with also account for the stability of the rotation law.

 If in the general equation (\ref{2ndeq}) we put $K$ and $D_{\mathrm{h}}$ equal to zero, the terms
 $B$ and $C$ in (\ref{BC}) are zero and
 we are left with the condition 
 \begin{equation}
 A= N^2_{\mathrm{ad}}+N^2_{\mu}+N^2_{{\Omega}-\delta v}  >  0 \; ,
 \label{solbmodif}
\end{equation}
 \noindent
 or explicitely
 \begin{eqnarray}
 N^2_{\mathrm{ad}}+N^2_{\mu}+\frac{1}{\varpi^3} \, \frac{d\left( \Omega^2 \, \varpi^4 \right) }{d\varpi} \, \sin \vartheta
 - \mathcal{R}i_{\mathrm{c}} \left(\frac{dv}{dr}\right)^2 > 0\, .
 \label{RiTayl}
 \end{eqnarray}
 \noindent
  This the
 modified form of the Solberg--H{\o}iland criterion for stability, accounting for the fact that 
  both the  distribution of $j$ and the energy excess in differential rotation
are considered simultaneously. 
The last term on the left hand side may reduce stability. Thus  in a differentially rotating star,
a situation considered as stable according to (\ref{solbh}) could in reality be unstable.

The Solberg--H{\o}iland criterion does not account for radiative losses, which would 
introduce the term $\Gamma/(\Gamma+1)$ in front of the first term
in (\ref{solbh}) and (\ref{RiTayl}). This would bring a dependence on the thermal
timescale. It neither account for
a viscosity due to turbulence, which would introduce a dependence on the viscous timescale. The Solberg--H{\o}iland instability essentially obeys the dynamical timescale.
The terms accounting for radiative losses and (turbulent) viscosity will appear below in the so--called GSF criterion.  In such a case, one would have a kind of semiconvective diffusion
like (\ref{dsc}), including in addition the effects of rotation. The characteristic timescales is determined by the thermal timescale rather than by the dynamical timescale.

\subsection{The Richardson criterion}  \label{sectri}

The usual expression of the Richardson criterion for shear instability in a star with shellular rotation is
\begin{eqnarray}
  N^2_{\mathrm{ad}} + N^2_{\mu} \, < \, \mathcal{R}i_{\mathrm{c}} \left( \frac{dv}{dr} \right)^2  \; .
 \label{Ri}
 \end{eqnarray}
 \noindent 
 It expresses that if the excess energy in the shear overcomes the restoring
 energy available in the stable temperature- and $\mu$-gradients, then
 mixing occurs.
 
 If we ignore thermal losses and horizontal turbulence,
 the modified form of the Richardson criterion is  identical to the above modified
 expression (\ref{RiTayl}) of the Solberg--H{\o}iland
 criterion, since the  shear effects have to be considered together with
 those regarding the stability of the angular momentum distribution.

 If account is given to the  thermal losses with the term $K$ and not to the horizontal turbulence,
 the stability relation  (\ref{2ndeq}) becomes,
 
\begin{eqnarray}
\left[N^2_{\mathrm{ad}}+N^2_{\mu}+N^2_{{\Omega}-\delta v}
\right]\, x+ \nonumber \\
\left[N^2_{\mu}\, K+ 
N^2_{{\mathrm{\Omega}}
-\delta v}
 \, K \right]  \,> \,0 \; .
\label{2ndeqK}
\end{eqnarray}
\noindent
 The corresponding  diffusion coefficient $D_{\mathrm{shear}}=2x$ is,
 \begin{eqnarray}
 D_{\mathrm{shear}} = 2 \,  K \, \left| \frac{N^2_{\mu} + N^2_{\Omega-\delta v} }
 {N^2_{\mathrm{ad}}+N^2_{\mu}+N^2_{\Omega-\delta v}} \, \right| \; .
 \label{Kmodif}
 \end{eqnarray}
 \noindent
 This is the diffusion coefficient for the effects of shear, with account of the $T$- and $\mu$-gradients, of the thermal losses and for the Rayleigh--Taylor instability. The complete 
 form of the stability criterion, including thermal losses and horizontal turbulence, is given by (\ref{2ndeq}).

 \subsection{The GSF criterion}

 The Goldreich--Schubert--Fricke (GSF) criterion \citep{GoldreichS67,Fricke68} considers the instability of a stellar rotation law subject to thermal diffusivity $K$
 and to viscosity $\nu$. Instability occurs for each of the two conditions,
 
  \begin{eqnarray}
  \frac{\nu}{K} \; N^2_{\mathrm{T, ad}}+N^2_{\Omega} \, < 0 
  \label{GSF1}\\  [2mm]
  \mathrm{and}  \quad \left|\varpi \frac{\partial \Omega^2}
  {\partial z} \right| \, >  \frac{\nu}{K} \; N^2_{\mathrm{T, ad}} \; , \label{GSF2}
  \end{eqnarray}
  
  \noindent
  The first inequality expresses the same idea as the Solberg--H{\o}iland criterion, with account of thermal diffusivity.
The second relation expresses the instability related to the differential rotation in the
 direction parallel to the rotation axis.

  The ratio $\nu/K$ can also be expressed in term of $\Gamma$
  as in Sect. \ref{thalother}, where we see a term $\Gamma/(\Gamma+1)$ in front of $N^2_{\mathrm{ad}}$. The viscosity is $\nu=(1/3) \, v \, \ell$
  and $\Gamma =v \,\ell /(6K)$. Formally one has the correspondence  $\Gamma = \nu /(2K)$ , the difference between a term like in (\ref{GSF1}) and  terms like $\Gamma/(\Gamma+1)$ comes from two facts: 
  
  -1.  In expressing $\Gamma$,  account is given to  the geometrical shape of the fluid element
  supposed to experience radiative losses, while in (\ref{GSF1}) this is not made. 
  
  - 2.  Also, as shown by \citet{Maeder95,maederlivre09}, in the general case of radiative losses one must consider
  the ratio $\Gamma /(\Gamma+1)$   in front of $N^2_{\mathrm{ad}}$. 
  It is only for an extreme conductivity $(\Gamma \rightarrow  0) $ of the medium that one may just write 
   $\Gamma$ or $\nu/K$. At the opposite, in the perfect adiabatic case, $\Gamma \rightarrow \infty$
   and the term in front of $N^2_{\mathrm{ad}}$ is 1.

 From these two remarks, one sees that the following writing of the standard GSF criterion  is more 
 appropriate,
  \begin{eqnarray}
  \left(  \frac{\Gamma}{\Gamma+1}\right) \, N^2_{\mathrm{T, ad}}+N^2_{\Omega} \, \sin \vartheta \, < 0 
  \label{GSF11}\\  [2mm]
  \mathrm{and}  \quad \left|\varpi \frac{\partial \Omega^2}
  {\partial z} \right| \, > \left(  \frac{\Gamma}{\Gamma+1}\right) \, N^2_{\mathrm{T, ad}} \; , \label{GSF22}
  \end{eqnarray}
\noindent
with the same notations as in Sect. \ref{thalother}.

The effect of $\mu$--gradient  can also be accounted for, as well as the diffusivity of chemical elements, due in particular to the horizontal turbulence. The instability is then  triple diffusive. In a rotating star, the horizontal turbulence
is a major diffusive process,
 its effects on the GSF instability have been studied \citep{MaederH10}. 
As for previous criteria, we also need here to account for
the excess energy present in the shears.
We do it in the same way as before in replacing $N^2_{\Omega}$ by
$N^2_{{\Omega}-\delta v}$. The first expression of the GSF instability
criterion is thus

\begin{eqnarray}
\left(\frac{\Gamma}{\Gamma+1}\right) N^2_{T, \, \mathrm{ad}} + \left(\frac{\Gamma_{\mu}}{\Gamma_{\mu}+1}\right) N^2_{\mu} +  N^2_{{\Omega}-\delta v}   < \, 0 \; . 
 \label{GSF1muri}
 \end{eqnarray}
 \noindent
 The second is 
\begin{eqnarray}
 \quad \left|\varpi \frac{\partial \Omega^2}
  {\partial z} \right|  > \left(  \frac{\Gamma}{\Gamma+1}\right)  N^2_{\mathrm{T, ad}} +
  \left(\frac{\Gamma_{\mu}}{\Gamma_{\mu}+1}\right) N^2_{\mu} - \frac{1}{4}
   \left(  \frac{dv}{dz} \right)^2 \; . 
   \label{GSF2muri}
  \end{eqnarray}
  
  These conditions are easier to realize  than the standard ones, 
  which means that the domain
  of the GSF instability is somehow extended by the
   account of the excess energy in the shear.
As shown previously \citep{MaederH10}, the GSF instability can be  locally  very
large in presupernova stages, however the zones over which the GSF instability
is acting are very narrow. It may be interesting to examine whether
the above expressions are changing the results. As a matter of fact, we see
that the first criterion is equivalent to our general stability equation
(\ref{2ndeq}).  The solution is given by the root of this quadratic equation.

 \section{Conclusions}

 We have considered simultaneously  various instabilities in differentially rotating stars and 
 given a general equation expressing the instability conditions with respect to an ensemble of effects,
 including horizontal turbulence.
 This introduces some coupling between the instabilities.
  We suggest that \emph{in rotating stars} some instabilities should never be dissociated. In particular the instability of rotation law expressed by Rayleigh
 criterion and the shear instability should always be considered simultaneously. 
Modeling the instabilities in 1 D models of stars with low or moderate rotation velocities is possible by considering the effects at the root
of $P_2(\cos \vartheta)=0$. In some future work now in preparation, we will  examine the consequences of the present approach on
 the mixing in stellar interiors.\\

\noindent
Acknowledgements: G.M., N.L. and C.C. acknowledge different supports from the Swiss National Fund.

\bibliographystyle{aa}
\bibliography{Thermopaper}

\begin{thebibliography}{30}
\expandafter\ifx\csname natexlab\endcsname\relax\def\natexlab#1{#1}\fi

\bibitem[{{Brown} {et~al.}(2012){Brown}, {Garaud}, \& {Stellmach}}]{Brown2012}
{Brown}, J., {Garaud}, P., \& {Stellmach}, S. 2012, ArXiv e-prints

\bibitem[{{Cantiello} \& {Langer}(2010)}]{CantielloLanger2010}
{Cantiello}, M. \& {Langer}, N. 2010, \aap, 521, A9

\bibitem[{{Chaboyer} \& {Zahn}(1992)}]{ChaboyerZ92}
{Chaboyer}, B. \& {Zahn}, J.-P. 1992, \aap, 253, 173

\bibitem[{{Chandrasekhar}(1961)}]{Chandrasekhar61}
{Chandrasekhar}, S. 1961, {Hydrodynamic and hydromagnetic stability}

\bibitem[{{Charbonnel} \& {Lagarde}(2010)}]{CL10}
{Charbonnel}, C. \& {Lagarde}, N. 2010, \aap, 522, A10

\bibitem[{{Charbonnel} \& {Zahn}(2007)}]{CZ07}
{Charbonnel}, C. \& {Zahn}, J.-P. 2007, \aap, 467, L15

\bibitem[{{Denissenkov}(2010)}]{D10}
{Denissenkov}, P.~A. 2010, \apj, 723, 563

\bibitem[{{Denissenkov} \& {Merryfield}(2011)}]{DM11}
{Denissenkov}, P.~A. \& {Merryfield}, W.~J. 2011, \apjl, 727, L8

\bibitem[{{Denissenkov} \& {Pinsonneault}(2008)}]{DenissenkovP08}
{Denissenkov}, P.~A. \& {Pinsonneault}, M. 2008, \apj, 684, 626

\bibitem[{{Endal} \& {Sofia}(1978)}]{EndalS78}
{Endal}, A.~S. \& {Sofia}, S. 1978, \apj, 220, 279

\bibitem[{{Fricke}(1968)}]{Fricke68}
{Fricke}, K. 1968, \zap, 68, 317

\bibitem[{{Goldreich} \& {Schubert}(1967)}]{GoldreichS67}
{Goldreich}, P. \& {Schubert}, G. 1967, \apj, 150, 571

\bibitem[{{Hirschi} \& {Maeder}(2010)}]{MaederH10}
{Hirschi}, R. \& {Maeder}, A. 2010, \aap, 519, A16

\bibitem[{{Kippenhahn} {et~al.}(1980){Kippenhahn}, {Ruschenplatt}, \&
  {Thomas}}]{KippenhahnRT80}
{Kippenhahn}, R., {Ruschenplatt}, G., \& {Thomas}, H.-C. 1980, \aap, 91, 175

\bibitem[{{Kippenhahn} \& {Thomas}(1970)}]{Kipp70}
{Kippenhahn}, R. \& {Thomas}, H.-C. 1970, in IAU Colloq. 4: Stellar Rotation,
  ed. A.~{Slettebak}, 20

\bibitem[{{Lagarde} {et~al.}(2012){Lagarde}, {Romano}, {Charbonnel}, {Tosi},
  {Chiappini}, \& {Matteucci}}]{Lagarde12}
{Lagarde}, N., {Romano}, D., {Charbonnel}, C., {et~al.} 2012, \aap, 542, A62

\bibitem[{{Langer} {et~al.}(1983){Langer}, {Fricke}, \&
  {Sugimoto}}]{LangerFS83}
{Langer}, N., {Fricke}, K.~J., \& {Sugimoto}, D. 1983, \aap, 126, 207

\bibitem[{{Maeder}(1995)}]{Maeder95}
{Maeder}, A. 1995, \aap, 299, 84

\bibitem[{{Maeder}(2003)}]{Maeder03Dh}
{Maeder}, A. 2003, \aap, 399, 263

\bibitem[{{Maeder}(2009)}]{maederlivre09}
{Maeder}, A. 2009, {Physics, Formation and Evolution of Rotating Stars}, ed.
  {Springer Berlin Heidelberg.}

\bibitem[{{Mathis} {et~al.}(2004){Mathis}, {Palacios}, \& {Zahn}}]{Mathis04Dh}
{Mathis}, S., {Palacios}, A., \& {Zahn}, J.-P. 2004, \aap, 425, 243

\bibitem[{{Meynet} \& {Maeder}(1997)}]{MMI}
{Meynet}, G. \& {Maeder}, A. 1997, \aap, 321, 465

\bibitem[{{Talon} \& {Zahn}(1997)}]{TZ97}
{Talon}, S. \& {Zahn}, J.-P. 1997, \aap, 317, 749

\bibitem[{{Th{\'e}ado} \& {Vauclair}(2012)}]{TheadoV12}
{Th{\'e}ado}, S. \& {Vauclair}, S. 2012, \apj, 744, 123

\bibitem[{{Traxler} {et~al.}(2011){Traxler}, {Garaud}, \&
  {Stellmach}}]{Traxler11}
{Traxler}, A., {Garaud}, P., \& {Stellmach}, S. 2011, \apjl, 728, L29

\bibitem[{{Ulrich}(1972)}]{Ulrich72}
{Ulrich}, R.~K. 1972, \apj, 172, 165

\bibitem[{{Vauclair} \& {Th{\'e}ado}(2012)}]{VauclairT12}
{Vauclair}, S. \& {Th{\'e}ado}, S. 2012, \apj, 753, 49

\bibitem[{{Wachlin} {et~al.}(2011){Wachlin}, {Miller Bertolami}, \&
  {Althaus}}]{Wachlin2011}
{Wachlin}, F.~C., {Miller Bertolami}, M.~M., \& {Althaus}, L.~G. 2011, \aap,
  533, A139

\bibitem[{{Wasiutynski}(1946)}]{Wasiutynski46}
{Wasiutynski}, J. 1946, Astrophysica Norvegica, 4, 1

\bibitem[{{Zahn}(1992)}]{Zahn92}
{Zahn}, J.-P. 1992, \aap, 265, 115

\end{thebibliography}

\end{document}